%% file: BPH-15-003_temp.tex
\pdfoutput=1

\documentclass[11pt,twoside,a4paper,cmspaper,final,collab]{cms-tdr}
\def\svnVersion{0613799-D}\def\svnDate{2020/04/08}\def\cmsCernNoTag{CERN-EP-2019-186}\def\cmsCernDate{\today}\def\cmsMessage{"Published in Physics Letters B as \href{http://dx.doi.org/10.1016/j.physletb.2020.135409}{\doi{10.1016/j.physletb.2020.135409}.}"}

\begin{document}\cmsNoteHeader{BPH-15-003}

\hyphenation{had-ron-i-za-tion}
\hyphenation{cal-or-i-me-ter}
\hyphenation{de-vices}
\newlength\cmsFigWidth
\ifthenelse{\boolean{cms@external}}{\setlength\cmsFigWidth{0.49\textwidth}}{\setlength\cmsFigWidth{0.9\textwidth}}
\ifthenelse{\boolean{cms@external}}{\providecommand{\cmsLeft}{upper\xspace}}{\providecommand{\cmsLeft}{left\xspace}}
\ifthenelse{\boolean{cms@external}}{\providecommand{\cmsRight}{lower\xspace}}{\providecommand{\cmsRight}{right\xspace}}
\providecommand{\cmsTable}[1]{\resizebox{\textwidth}{!}{#1}}
\newlength\cmsTabSkip\setlength{\cmsTabSkip}{1ex}

\cmsNoteHeader{BPH-15-003} 
\title{Study of \texorpdfstring{$\PJGy$}{J/psi} meson production inside jets in $\Pp\Pp$ collisions at \texorpdfstring{$\sqrt{s} = 8\TeV$}{sqrt(s)= 8 TeV}}

\date{\today}

\abstract{
A study of the production of prompt $\PJGy$ mesons contained in jets in proton-proton collisions at $\sqrt{s} = 8\TeV$ is presented.  The analysis is based on data corresponding to an integrated luminosity of 19.1\fbinv collected with the CMS detector at the LHC.  For events with at least one observed jet, the angular separation between the $\PJGy$ meson and the jet is used to test whether the $\PJGy$ meson is part of the jet. The analysis shows that most prompt $\PJGy$ mesons having energy above 15\GeV and rapidity $\abs{y}<1$ are contained in jets with pseudorapidity $\abs{\eta_{\text{jet}}} <1$.  The differential distributions of the probability to have a $\PJGy$ meson contained in a jet as a function of jet energy for a fixed $\PJGy$ energy fraction are compared to a theoretical model using the fragmenting jet function approach.   The data agree best with fragmenting jet function calculations that use a long-distance matrix element parameter set in which prompt $\PJGy$ mesons are predicted to be unpolarized.  This technique demonstrates a new way to test predictions for prompt $\PJGy$ production using nonrelativistic quantum chromodynamics. 
}

\hypersetup{%
pdfauthor={CMS Collaboration},%
pdftitle={Study of J/psi meson production inside jets in pp collisions at sqrt(s) = 8 TeV},%
pdfsubject={CMS},%
pdfkeywords={CMS, J/psi Mesons, jets}}

\maketitle

\section{Introduction}

The mechanism for producing $\PJGy$ mesons as bound states of charm quark pairs ($\PQc\PAQc$) in hadronic collisions has been under intensive experimental and theoretical study since the 1974 discovery of the  $\PJGy$ meson in proton-nucleon collisions~\cite{ting} and in $\Pep\Pem$ annihilations~\cite{richter}.  The early theoretical descriptions of the hadronic production process considering only color-singlet production~\cite{baier,lans1} were at odds with the differential cross section measurements as a function of the $\PJGy$ transverse momentum $\PT^{\PJGy}$ made by experimenters at the Fermilab Tevatron~\cite{cdf2} for $\PT^{\PJGy}>6\GeV$.  A new theoretical approach, nonrelativistic quantum chromodynamics (NRQCD), was used to address the problem~\cite{bbl,NRQCD1,NRQCD2}. The NRQCD model includes both color-singlet and color-octet amplitudes for the $\PQc\PAQc$ system that ultimately produces the $\PJGy$ meson.  It proved to be capable of explaining the cross section data, using phenomenological parameters called long-distance matrix elements (LDMEs) that are adjusted to describe $\PJGy$ meson production data.  Within the NRQCD factorization assumption, the LDME parameters are process independent.  However, each determination of an LDME set can choose a specific collection of $\PJGy$ meson production data and $\PJGy$ meson kinematic requirements.  Furthermore, different LDME sets that describe the production data may have different predictions for the $\PJGy$ meson polarization~\cite{faccioli}.   Experiments~\cite{prl99,cmspol} have shown that the prompt $\PJGy$ meson polarization at large $\PT^{\PJGy}$ ($>$12\GeV) is small.  Recent NRQCD studies extend the range of experimental input to include low-$\PT$ data and attempt to make global fits to the full set of charmonium information.  A review of these studies can be found in Ref.~\cite{jpl}. 

A remaining theoretical problem is to determine the mechanism by which a $\PQc\PAQc$ system in an angular momentum state and quark color configuration $^{2S+1}L_{J}^{n}$ hadronizes into a $\PJGy$ meson.  Here, $S$, $L$, and $J$ are the spin, orbital, and total angular momentum quantum numbers of the $\PQc\PAQc$ system.  Its color state is labeled by $n$, with $n = 1$ or 8 referring to a color-singlet or color-octet configuration, respectively.  The $\PJGy$ meson has $J=S=1$ and $n=1$.  The analysis described in this Letter combines the experimental measurement of $\PJGy$ mesons contained in jets with a theoretical approach based on the fragmenting jet function (FJF) model~\cite{ira}.   The FJF model postulates that the $\PQc\PAQc$ pair is not produced directly in the hard scattering, but is a fragmentation product of a high-$\PT$ jet. The model uses the methodology of NRQCD to compute the cross section contributions for all relevant $^{2S+1}L_{J}^{n}$ terms.  Each cross section term has a characteristic relation between the jet energy $E_{\text{jet}}$ and its fraction carried by the $\PJGy$ meson: $z = E_{\PJGy}/E_{\text{jet}}$.

A study of $\PJGy$ mesons contained in jets in the rapidity region $y_{ \PJGy}>2$, dominated by charm fragmentation for large $z$, has been reported by the LHCb Collaboration~\cite{lhcb}.  The LHCb analysis, which measured the $z$ distribution integrated over jet energy, does not have the sensitivity to LDME parameter sets that characterizes this analysis.

 The data for this analysis were collected by the CMS detector in proton-proton ($\Pp\Pp$) collisions from the CERN LHC, corresponding to an integrated luminosity of 19.1\fbinv at $\sqrt{s} = 8\TeV$.  It is the first experimental study of prompt $\PJGy$ mesons contained in jets produced in the gluon-dominated central rapidity region, where the FJF theory for gluonic jet fragmentation applies.

\section{Theoretical framework}

The hadronization process is nonperturbative. It is handled in the FJF approach by an NRQCD expansion of the fragmentation function for a jet initially produced in a hard scattering at high energy.  The observables are $E_{\text{jet}}$ and $z$.  Following Ref.~\cite{ira}, the differential cross section for dijet production, with one jet fragmenting to a $\PJGy$ meson, can be written symbolically as 
\begin{linenomath}
\ifthenelse{\boolean{cms@external}}
{
\begin{multline}
 \frac{\rd^2 \sigma(E_{\text{jet}};z)}{\rd E_{\text{jet}}\,\rd z} = \sum_{A,\,B,\,i,\,j} f_{A/p} \  f_{B/p} \ \rd\sigma_{ABij}(\PQc\PAQc X, n, \mathcal{J}_j)\\
  \otimes \ \mathcal{F}_S \otimes \ \mathcal{G}_i^{\PJGy}(E_{\text{jet}},z|R,\mu).~\label{FJF}
\end{multline}
}
{
\begin{equation}
 \frac{\rd^2 \sigma(E_{\text{jet}};z)}{\rd E_{\text{jet}}\,\rd z} = \sum_{A,\,B,\,i,\,j} f_{A/p} \  f_{B/p} \ \rd\sigma_{ABij}(\PQc\PAQc X, n, \mathcal{J}_j)   \otimes \ \mathcal{F}_S \otimes \ \mathcal{G}_i^{\PJGy}(E_{\text{jet}},z|R,\mu).~\label{FJF}
\end{equation}
}
\end{linenomath}
In this expression, $A$ and $B$ are the partons in the colliding protons with fractional flavor content $f_{A/p}, f_{B/p}$, respectively, while $i$ and $j$ are the outgoing partons.  The symbolic hard-scattering cross section $\rd\sigma_{ABij}(c \overline{c} X, n, \mathcal{J}_j)$ produces the fragmenting jet from outgoing parton $i$ and the recoil jet $\mathcal{J}_j$ from outgoing parton $j$.  The fragmenting jet produces a $\PQc\PAQc$ system characterized by $S$, $L$, $J$, and $n$ quantum numbers, plus an inclusive hadronic state $X$ that represents the remainder of the jet.  The function $\mathcal{F}_S$ controls the evolution of the fragmenting system down to the energy scale $\mu$ equal to the mass of the $\PQc\PAQc$ system, to allow the development of jet structure from soft gluons.  The nonperturbative fragmentation of the $\PQc\PAQc$ system into the observed $\PJGy$ meson is described by the function $\mathcal{G}_i^{\PJGy}(E_{\text{jet}},z | R,\mu)$, where $E_{\text{jet}}$  is determined in a cone of angular radius $R$.

The type of parton $i$ that produces the fragmenting jet, and ultimately the $\PJGy$ meson, depends on the jet rapidity region.  In the central rapidity region covered by this analysis, gluon fragmentation dominates~\cite{Pumplin}.  The FJF expression for $\mathcal{G}_i^{\PJGy}$ sums over all contributing partons, but the light flavor contributions are suppressed and can be neglected.  In Ref.~\cite{ira}, the small central charm quark fragmentation contribution was absorbed into the $^{3}S_1^{1}$ contribution to gluon fragmentation, so $\mathcal{G}^{\PJGy}$ in this Letter represents only gluon fragmentation. 

In Ref.~\cite{ira2}, the authors updated the work of Ref.~\cite{ira} to make an explicit computation of the perturbative dijet double-differential cross section, followed by the fragmentation of one of the jets to a $\PJGy$ meson.  They integrated over the kinematic variables of the second jet to give an FJF expression for the absolute differential cross section to produce a jet of energy $E_{\text{jet}}$ that fragments into a $\PJGy$ meson carrying energy fraction $z$ of the parent jet energy along with the remaining fragments.  In the NRQCD decomposition of $\mathcal{G}^{\PJGy}$ for central $\PJGy$ meson hadroproduction with $\PT > 10\GeV$, four FJF terms are relevant: $^{3}S_1^{1}, \ ^{1}S_0^{8}, \ ^{3}S_1^{8}, \ \text{and} \ ^{3}P_J^{8}$.  Only the $^{1}S_0^{8}$ term has all angular momenta equal to zero in the $\PQc\PAQc$ rest frame.  If this NRQCD term were to dominate the jet fragmentation process, then the $\PJGy$ meson would be produced unpolarized.

\section{The CMS detector} 

The central feature of the CMS apparatus is a superconducting solenoid of 6\unit{m} internal diameter, providing a magnetic field of 3.8\unit{T}. Within the solenoid volume are a silicon pixel and strip tracker, a lead tungstate crystal electromagnetic calorimeter, and a brass and scintillator hadron calorimeter, each composed of a barrel and two endcap sections.  When combining information from the entire detector, the jet energy resolution amounts typically to 15\% at 10\GeV and 8\% at 100\GeV~\cite{jetart}.  Muons are detected in gas-ionization chambers embedded in the steel flux-return yoke outside the solenoid, covering the pseudorapidity range $\abs{\eta}<2.4$.  The silicon tracker measures charged particles within the pseudorapidity range $\abs{\eta}<2.5$. It consists of 1440 silicon pixel and 15\,148 silicon strip detector modules. For nonisolated particles of $1<\pt<10\GeV$ and $\abs{\eta}<1.4$, the track resolutions are typically 1.5\% in \pt and 25--90 (45--150)\mum in the transverse (longitudinal) impact parameter \cite{TRK-11-001}.   Matching muons to tracks measured in the silicon tracker results in a relative transverse momentum resolution, for muons with $20<\pt<100$\GeV, of 1.3--2.0\% in the barrel~\cite{Chatrchyan:2012xi}.  Events of interest are selected using a two-tiered trigger system~\cite{Khachatryan:2016bia}. The first level, composed of custom hardware processors, uses information from the calorimeters and muon detectors to select events at a rate of around 100\unit{kHz} within a fixed time interval of less than 4\mus. The second level, known as the high-level trigger (HLT), consists of a farm of processors running a version of the full event reconstruction software optimized for fast processing.  This reduces the event rate to around 1\unit{kHz} before data storage.  A more detailed description of the CMS detector, together with a definition of the coordinate system used and the relevant kinematic variables, can be found in Ref.~\cite{cmsdet}.

\section{Event selection and background subtraction} \label{anal}
  
The experimental methods follow those used by previous CMS analyses of inclusive $\PJGy$ and $\Upsilon(n$S) production at $\sqrt{s} = 7\TeV$~\cite{j1,y1,Chatrchyan:2011kc,jpsixs,upsixs}.  The event selection is based on a dimuon trigger involving the silicon tracker and muon systems. The trigger requires two oppositely charged muons with rapidity of the dimuon system $\abs{y}<1.25$ and its invariant mass range $2.7<m_{\Pgm\Pgm}<3.5\GeV$.  The three-dimensional fit to the dimuon vertex must have a $\chi^{2}$ probability (the $p$-value of the $\chi^2$ returned by the fit) ${>}0.5\%$.  Only dimuon pairs in which the muons bend away from each other in the magnetic field are used to allow a precise dimuon efficiency determination.  The dimuon \PT trigger threshold varied from 5 to 9\GeV during the data-taking period.  The primary event vertex is defined as the one with the largest summed \PT of its associated tracks.

The offline selection requires a dimuon pair with $\pt >10\GeV$, $\abs{y}<1$, energy $E>15\GeV$, and vertex fit $\chi^{2}$ probability ${>}1\%$.  In order to guarantee agreement to within 3\% between the single-muon efficiencies from control samples and from simulation, each muon must have $\PT^{\Pgm}>6\GeV$ and $\abs{\eta_{\Pgm}}<2.1$, or $\PT^{\Pgm}>5\GeV$ and $\abs{\eta_{\Pgm}}<0.8$.  The muon candidate must satisfy the CMS ``tight'' muon quality requirements on the number of tracker hits, the muon track fit quality, and the distance along the beam line from the primary event vertex~\cite{Chatrchyan:2012xi}.  No muon isolation requirements are applied, because we look for $\PJGy$ + jet associations.  The $\PJGy$ signal invariant mass range is $2.95<m_{\Pgm\Pgm}<3.20\GeV$.  After the data selection, we observe at most one $\PJGy$ candidate per event.   

The trigger does not use any information about jets in the event.  Jets are reconstructed from particle-flow objects~\cite{jetres}, using an anti-\kt algorithm with a distance parameter of 0.5~\cite{kt}, as implemented in the \textsc{FastJet} package~\cite{fastjet}.   The jet response has been corrected to the particle level~\cite{jetart}.  Although the $\PJGy$ candidate is not a particle-flow object, its decay muons are.  This does not exclude jets that consist only of a $\PJGy$ meson.  However, such jets constitute less than $10^{-4}$ of this sample and do not affect the results presented here.  The jet properties include the energy $E_{\text{jet}}$, the transverse momentum magnitude $\PT^{\text{jet}}$, the number of constituents, and the number of included muons.  Each bunch crossing in the data produces, on average, 14 reconstructed $\Pp\Pp$ vertices, corresponding to 21 interactions per bunch crossing.  The extra interactions produce so-called pileup distortions, which are corrected using the procedure described in Ref.~\cite{jetart}.  For this analysis, the jet selection requirements are $\PT^{\text{jet}}>25\GeV$ and $\abs{\eta_{\text{jet}}}<1$.   

The $\PJGy$ event candidates are classified as prompt, nonprompt, or combinatorial.  Nonprompt events include those $\PJGy$ mesons that come from decays of $b$ hadrons.  Combinatorial candidates are accidental pairings of an identified $\PGmp$ and a $\PGmm$ such that the dimuon invariant mass falls within the signal mass interval.  The nonprompt background is strongly reduced by applying a selection on the variable $\Sigma_{\text{TD}}$, which is the sum of the squares of the significance (in units of standard deviations) of the transverse distance of closest approach of each muon track to the primary vertex. The $\Sigma_{\text{TD}}$ distribution has a sharp peak near zero from prompt events and a long tail at larger $\Sigma_{\text{TD}}$ from nonprompt sources, which we fit with an exponential function.  From a prompt $\PJGy$ meson Monte Carlo (MC) sample, we find that ${>}99\%$ of the events have $\Sigma_{\text{TD}}<10$.  The simulated $\Sigma_{TD}$ shape agrees with that in data for this region, so we require $\Sigma_{\text{TD}}<10$ to define the prompt dimuon events.  In the $\PJGy$ data, the exponential function that describes the nonprompt background is extrapolated into the range $\Sigma_{TD} < 10$ to estimate the fraction of nonprompt events in the prompt signal mass range.  This is ($5.7\pm0.1$)\%.  The events in the prompt signal mass range also contain combinatorial background, which is determined by interpolating the $m_{\PGm\PGm}$ low (2.70--2.90\GeV) and high (3.25--3.50\GeV) sideband regions. We find that the combinatorial background fraction in the prompt signal mass range is ($1.4\pm0.2$)\%.  The quoted uncertainties in the backgrounds are statistical only.   All distributions shown in this Letter have the nonprompt and combinatorial backgrounds subtracted.   After background subtraction, there are $1.63 \times 10^6$ prompt $\PJGy$ meson candidates.

\section{Association of jets and \texorpdfstring{\PJGy}{J/psi} mesons }\label{assn}

\begin{figure}[h]
\centering
 \includegraphics[width=0.49\textwidth]{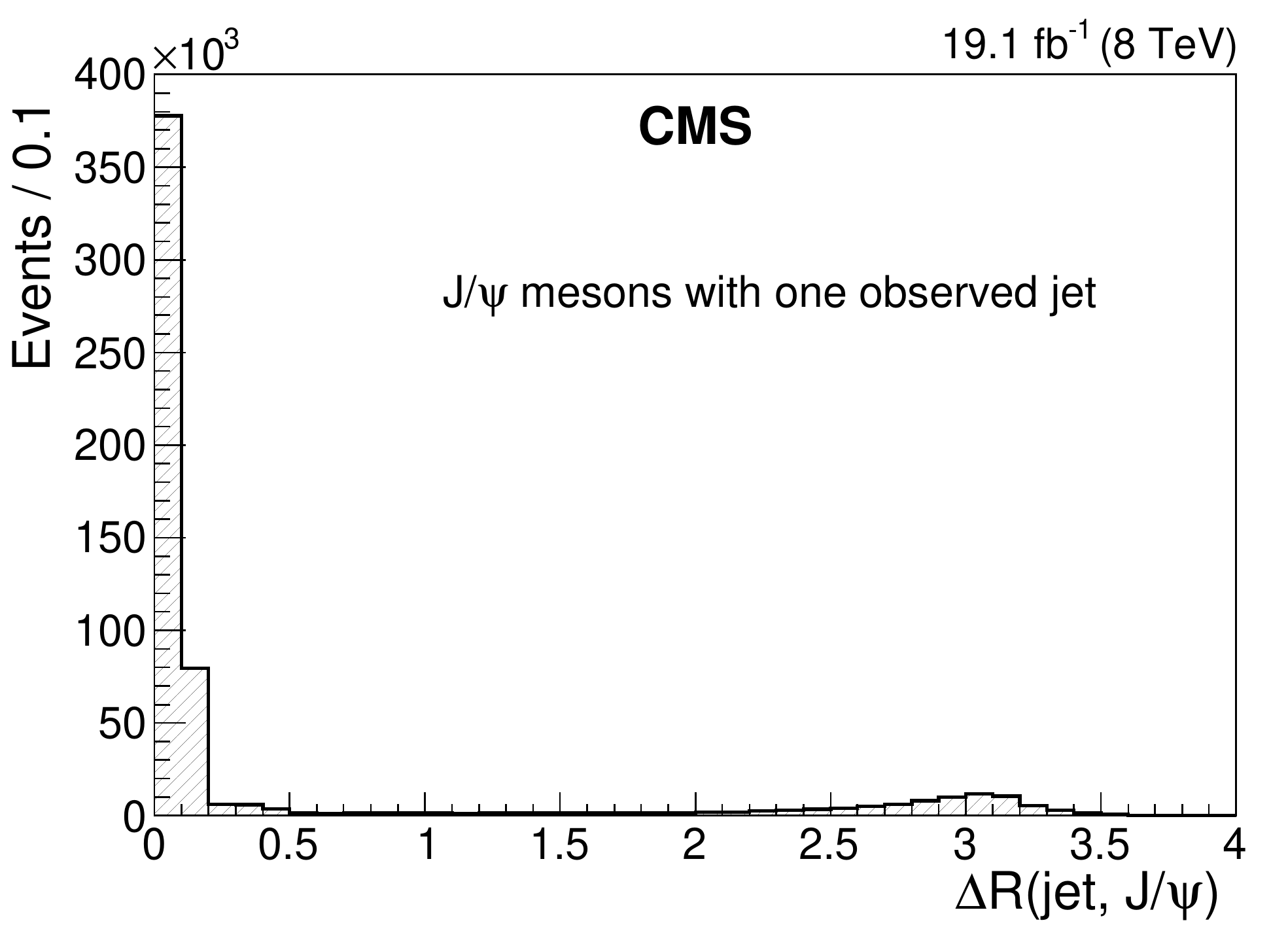} \includegraphics[width=0.49\textwidth]{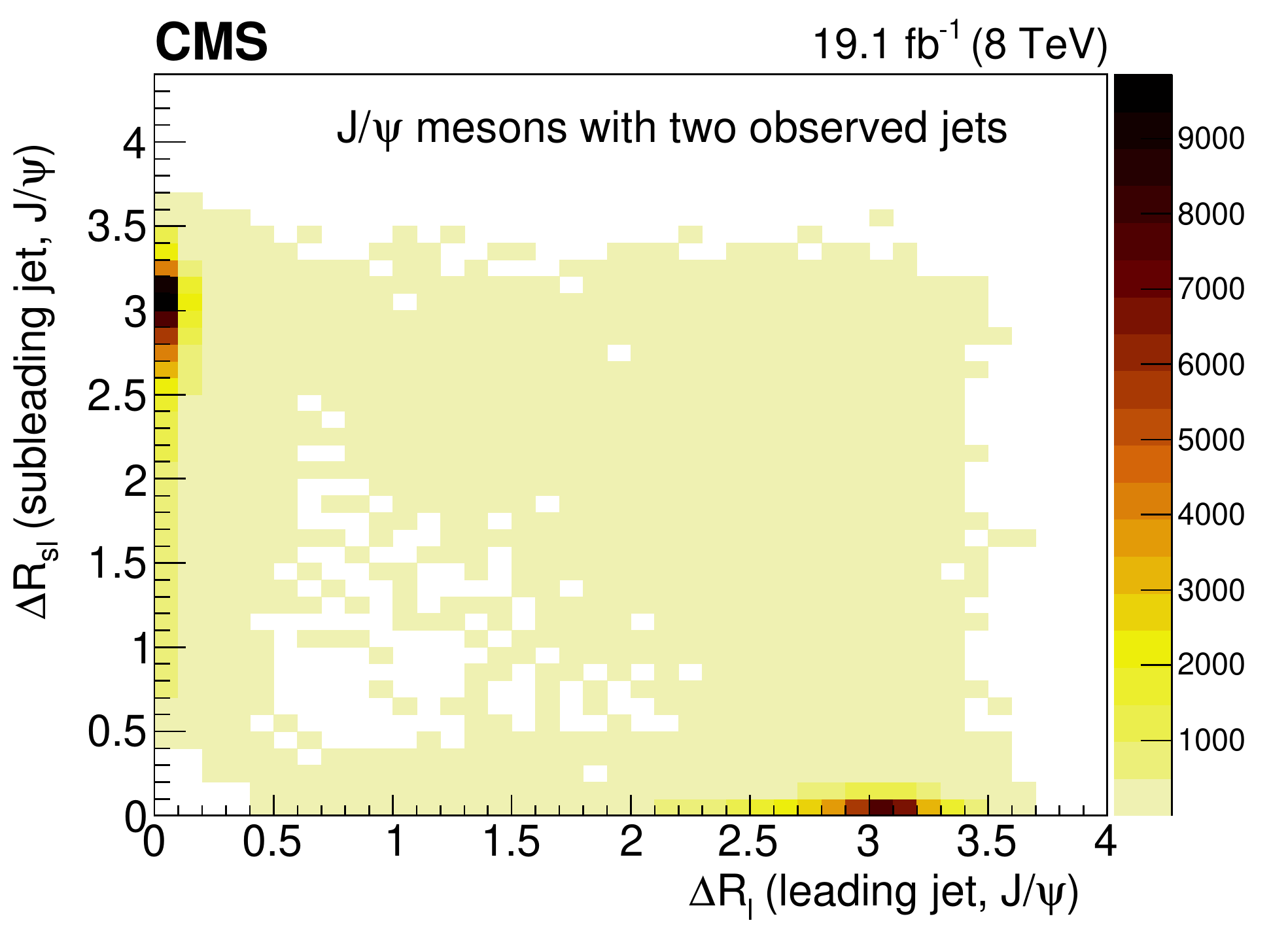}
\caption{ The distributions of (\cmsLeft) $\DR$(jet, $\PJGy$) for one-jet events and (\cmsRight) $\DR_l$(leading jet, $\PJGy$) vs. $\DR_{sl}$(subleading jet, $\PJGy$) for two-jet events\label{jetfrac1}}
\end{figure}

The analysis makes no restriction on the number of jets that pass the jet selection requirements, which we term ``observed jets''.  For jet requirements $\PT^{\text{jet}} > 25\GeV$ and $\abs{\eta_{\text{jet}}}< 1$, the fractions of $\PJGy$ meson events that have 0, 1, 2, or 3 observed jets are ($55.12\pm0.06$)\%, ($34.03\pm0.05$)\%, ($9.58\pm0.02$)\%, and ($1.27\pm0.08$)\%, respectively, where the uncertainties are statistical only.  For events with at least one observed jet, the association of a $\PJGy$ meson with a jet is made using the angular separation \ $\DR=\sqrt{\smash[b]{( \eta_{\text{jet}} -\eta_{\PGm\PGm} )^2 + ( \phi_{\text{jet}}-\phi_{\PGm\PGm} )^2}}$.  Here, $\eta_{\text{jet}} \ (\eta_{\PGm\PGm})$ and $\phi_{\text{jet}} \ (\phi_{\PGm\PGm})$ are the pseudorapidity and azimuthal angle (modulo $\pi$), respectively, of the jet (dimuon) direction.  The $\DR$ distribution for the best-matched jet is sharply peaked at zero, as seen for events with one observed jet in Fig.~\ref{jetfrac1} (\cmsLeft).  The $\PJGy$ meson and the jet are defined as associated if $\DR<0.5$.  Furthermore, if both decay muons from the $\PJGy$ meson are among the objects that comprise the jet, we say that the $\PJGy$ meson is a constituent of the jet. 

When there are two observed jets in the event, further evidence that $\PJGy$ meson production comes primarily from jets is shown in Fig.~\ref{jetfrac1} (\cmsRight).  This plot shows $\DR$ for the $\PJGy$ meson with respect to each observed jet in two-jet events.  The higher-energy (leading) jet has $\DR_{\text{l}}$, the lower-energy (subleading) one $\DR_{\text{sl}}$. Note that the energy labels here play no role in the analysis; the jets need only to pass the jet $\PT$ and $\abs{\eta_{\text{jet}}}$ requirements given above.  The $\PJGy$ meson is not required to come from either jet.  The clusters of events in Fig.~\ref{jetfrac1} (\cmsRight), near ($\DR_{l}, \ \DR_{sl}) = (0, \pi)$ and ($\pi$, 0), show that ($94.1\pm0.1$)\% of the time, the $\PJGy$ meson is a constituent of one of the two jets in the event.  In events with a $\PJGy$ meson and two jets, the mean and RMS deviation of the distribution of the number of jet constituents, charged and neutral, for the fragmenting jet ($25\pm8$) and the recoil jet ($29\pm8$) are similar.  The shapes of the jet energy spectra for the jet containing the $\PJGy$ meson and the recoil jet are indistinguishable.  The difference in the probability for the $\PJGy$ meson to be a jet constituent in the one- and two-jet cases, along with a discussion of the small excess for $2.4< \DR <3.5$ in Fig.~\ref{jetfrac1} (\cmsLeft), will be addressed in Section~\ref{prod}.

\section{Experimental application of the FJF approach}\label{exanal}

The authors of Refs.~\cite{ira,ira2} emphasize that experimental sensitivity to the FJF terms in jet fragmentation comes from measuring the jet energy dependence of the function $\mathcal{G}$ in Eq.~(\ref{FJF}) at fixed $z$.  In the FJF framework, the dependence of the fragmenting jet differential cross section on the $\PJGy$ properties comes solely through the $z$ variable.  Integrating Eq.~(\ref{FJF}) over $z$ gives the single-jet differential cross section for the production of $\PJGy$ mesons contained in jets, as a function of $E_{\text{jet}}$.  This is used as a normalization term in Ref.~\cite{ira2}, where the differential cross section for a jet to fragment to a $\PJGy$ meson with the energy fraction $z$ is calculated for jets having $\PT^{\text{jet}} >25\GeV$ and pseudorapidity $\abs{\eta_{\text{jet}}}< 1.2$.  The resulting $\PJGy$ meson is required to have energy above 15\GeV and rapidity $\abs{y_{ \PJGy}}<1$.  The jet fragmentation cross section is normalized by integrating over the $z$ range 0.3--0.8.  The authors showed that the jet energy dependence of the normalized FJF terms is insensitive to the exact $z$ range used.  At a fixed $z$ value, called $z_1$, the ratio of the fragmenting jet differential cross section due to a single FJF term $i$ to the sum of the cross section integrals for $0.3< z<0.8$ for all FJF terms is termed $(\rd \tilde{\sigma_i}/\rd E_{\text{jet}} \ \rd z)|_{z_1}$ in Ref.~\cite{ira2}.  The sum of this ratio over all four FJF terms is denoted as $(\rd \tilde{\sigma}/\rd E_{\text{jet}} \ \rd z)|_{z_1}$.  For a given LDME parameter set, each of the four FJF terms is different.  Also, changing the LDME parameter set changes the FJF predictions for the four terms.

The authors of Ref~\cite{ira2} cite next-to-leading order (NLO) calculations~\cite{bodwin,nlo1,nlo2,nlo3} to argue that the $\PT^{\PJGy}$ range for the three $z_1$ values used in this analysis is large enough that the $^3S_1^{1}$ NRQCD term cannot contribute to the sum.  Therefore, in computing $(d \tilde{\sigma}/d E_{\text{jet}} \ d z)|_{z_1}$ to compare to these data, only the three color-octet terms are included.  However, in the low-$z$ region included in the normalizing integral, the $^3S_1^{1}$ NRQCD term can play a role and is included in their calculation for $0.3< z<0.8$.  

The experimental proxy for $(\rd \tilde{\sigma}/\rd E_{\text{jet}} \ \rd z)|_{z_1}$, evaluated for a jet energy bin centered at $E_{\textrm{c}}$, is called $\Xi(E_{\textrm{c}};z_1)$:
\begin{linenomath}
\begin{equation}
\Xi(E_{\textrm{c}};z_1) \equiv \frac{ N(E_{\textrm{c}};z_1)}{\int_{0.3}^{0.8} N(E_{\textrm{c}};z) \ \rd z} \ ,\label{xieq}
\end{equation} 
\end{linenomath}
where $N(E_{\textrm{c}};z_1)$ is the number of events having a $\PJGy$ meson contained in a jet for a $z$ interval $\Delta_z$ centered on $z_1$ in that $E_{\text{jet}}$ bin.  The number of events is weighted to correct for the $\PJGy$ meson efficiency and acceptance, as described in Section~\ref{eff}, as well as corrected for jet efficiency and jet energy resolution, as described in Section~\ref{unf}.  We use a $z$ interval $\Delta_z = \pm$~0.025 around $z_1$, which is small enough to be insensitive to $z$ variations in $\Xi$ and large enough to provide a reasonable number of events in each $E_{\text{jet}}$ bin.

\section{Efficiency corrections for \texorpdfstring{\PJGy}{J/psi} mesons}\label{eff}

Measuring the properties of events when a $\PJGy$ meson is a jet constituent requires an event-by-event $\PJGy$ meson efficiency correction. Each entry in the signal or background event distributions has an event weight, defined as 1/$\epsilon_{\PJGy}$.  The dimuon acceptance times efficiency $\epsilon_{\PJGy}$ is determined using a simulated sample of unpolarized $\PJGy$ meson events, uniformly distributed in 1\GeV wide $\PT$ bins and uniformly distributed over $\abs{y_{\PJGy}} < 1.5$.  Only the $\PJGy$ meson is simulated; studies~\cite{jpsixs,upsixs} show that using a complete $\PYTHIA$~\cite{pythia} event simulation does not change the efficiency results.  The $\PJGy \rightarrow \PGmp\PGmm$ decay is simulated using $\EVTGEN$~\cite{evtgen}; radiative effects are treated by $\PHOTOS$~\cite{PHOTOS1}; and the detector response to the two muons is simulated using the $\GEANTfour$-based~\cite{geant4} CMS simulation program.  The simulated $\PJGy$ meson must pass the quality requirements listed in Section~\ref{anal}.  The total efficiency $\epsilon_{\PJGy}$ varies with the rapidity and transverse momentum of the $\PJGy$ meson because the muon reconstruction, dimuon vertex reconstruction, and dimuon trigger efficiencies depend on these variables. There is also an HLT trigger inefficiency if two muons in the event have a small angular separation.  This is also taken from simulation and checked against data taken using a single-muon trigger.

\section{Jet energy corrections and unfolding}\label{unf} 

A crucial part of the analysis is measuring the energy of the jet that contains the $\PJGy$ meson.  To test whether there might be an influence on the jet energy distribution due to the presence of the $\PJGy$ meson, we study the two-jet events shown in Fig.~\ref{jetfrac1} (\cmsRight).  The energy distributions of the fragmenting jet and the recoil jet are compared for $0.3< z <0.8$ and for $z$ ranges of 0.40--0.45, 0.50--0.55, and 0.60--0.65.  The shapes of the measured energy distributions of the recoil and fragmenting jets for each sample are indistinguishable.  There is no evidence that having a $\PJGy$ meson as a constituent affects the jet energy distribution.

The jet energy distributions are compared to the FJF model predictions in bins of jet energy.  Experimentally, the jet energy bin width $\Delta E_{\text{jet}}$ is constrained by the finite jet energy resolution of the CMS apparatus, which must be unfolded.  We use $\Delta E_{\text{jet}}=8$\GeV.  The D'Agostini unfolding method~\cite{DAgostini} from the \textsc{RooUnfold} package~\cite{adye} is used to extract the unsmeared $\Xi$ distribution.  The procedure uses an input generator-level jet energy distribution (truth distribution) derived from \textsc{pythia} light-quark or gluon jets.  Simulation shows that for measured jet energy $E_{\text{jet}}>44$\GeV, the jet reconstruction efficiency exceeds 98.5\% and is consistent with being energy independent.  Thus, 44\GeV is the lowest jet energy considered in the unfolding procedure.  The unfolding procedure uses the CMS jet energy resolution and jet finding efficiency~\cite{jetres} to determine the unfolded jet energy matrix and the MISS matrix.  The latter is filled for events that fail the jet efficiency test or fall outside the unfolded jet energy window 44--140\GeV.  The method was validated using several different simulated jet energy input truth distributions, including a fit to the \textsc{pythia} shape using the sum of exponentials and the raw data itself in a bootstrap approach.  There was no change in the unfolded distributions that exceeded $\sigma_{stat}$/4 for any jet energy bin.  Based on unfolding studies in simulation that used three to six iterations, we found that four unfolding iterations gave stable matches to the simulation events and showed no sensitivity to different choices for the input truth distribution.  Based on the simulation results, the unfolded jet energy range is $56<E_{\text{jet}}<120$\GeV.  This range is stable when the input distribution is changed.  Henceforth, $E_{\text{jet}}$ will refer to the unfolded quantity, unless otherwise noted.

The unfolded jet energy distributions for the $\Xi(E;z)$ functions have bin-to-bin correlations that affect the statistical uncertainty in $\Xi$ for each jet energy bin.  The statistical uncertainties are evaluated by repeating the unfolding procedure 250 times, forming the covariance matrix, and determining the uncertainty for each jet energy bin.  The statistical uncertainties computed by this procedure are 0.02 to 0.06\%.  The unfolding in $z$ is dominated by the $E_{\text{jet}}$ resolution.  The changes in $z$ from the unfolding procedure for the region of interest (0.40--0.65) are less than 0.01 in $z$.  Therefore, the measured $z$ values are used in the $\Xi(E_{\text{jet}}; z)$ determinations.

\section{Systematic uncertainties}\label{syst}
The systematic uncertainties arise from the determination of the event weight, based on the $\PJGy$ meson and muon properties, and from a bias in the $\PJGy$-jet association, discussed below.  The systematic uncertainty in the jet energy scale is small compared to the jet energy resolution used in the unfolding.  Varying the jet energy by the jet energy scale systematic ($<2.2\%$) uncertainty before the unfolding made no change in the $\Xi$ results. 

The CMS studies at $\sqrt{s} = 8\TeV$ using a tag-and-probe method~\cite{j1,y1} show that, for the offline requirements used in this analysis, the ratio of the single-muon efficiency in data and MC simulation is consistent with unity within ${<}3\%$, independent of $\PT^{\mu}$~\cite{ilse}.  The tracking efficiency in data and simulation agree to within 1\% per track.  The dimuon vertex and trigger simulation also have 1\% systematic uncertainties.  The dimuon HLT trigger inefficiency varies with $\PT^{\PJGy}$ in the range 4.5--7.5\%.  For the few dimuons with $\PT>60$\GeV, it can go up to 15\%. The difference between unity (no loss) and the simulated HLT trigger efficiency is assigned as the HLT systematic uncertainty for each event. All of the above systematic uncertainties are added in quadrature to determine the total systematic uncertainty in the weight for each event.  To estimate the impact of the weight systematic uncertainty on the $\Xi(E_{\text{jet}}; z_1)$ function, two additional $\Xi(E_{\text{jet}}; z_1)$ functions are made for each $z_1$.  One uses distributions in which the weight for each event is raised by one standard deviation; in the other, the event weight is lowered by one standard deviation.  The shifted $\Xi(E_{\text{jet}}, z_1)$ values are compared to the unshifted value in each energy bin.  The relative systematic uncertainty in the event weight ranges from 0.2 to 0.9\% of the standard-weight $\Xi(E_{\text{jet}}; z_1)$ values.

In addition, there is a selection bias in the $\PJGy$ meson and jet association that disfavors the configuration when the difference $\eta_{\text{jet}}-\eta_{\PJGy}$ has the opposite sign to $\eta_{\text{jet}}$.  The bias originates from events that are lost in the section on $\abs{\eta_{\text{jet}}-\eta_{\PJGy}}$ and is evaluated from data.  The number of events per $E_{\text{jet}}$ bin in the biased region is rescaled to match the yield in the unbiased region.  Half of the difference between the measured and corrected number of events in each $E_{\text{jet}}$ bin is assigned as its bias systematic uncertainty.  The weight and bias systematic uncertainties are added in quadrature to obtain the systematic uncertainty in $\Xi(E_{\text{jet}}; z_1)$, which ranges from 0.3 to 1.0\%.  These uncertainties are then added in quadrature with the uncertainty in the unfolding procedure discussed in the previous section.

\section{The FJF predictions of the jet energy spectrum}\label{FJFanal}

In this analysis, we use three $z_1$ values: 0.425, 0.525, and 0.625.  These are the centers of three nonoverlapping $z$ subregions with $\Delta z = 0.05$ from the measurement region $0.3<z<0.8$.  In these three $z$ regions, the FJF terms have different jet energy distributions for a given LDME parameter set.  The authors of Ref.~\cite{ira2} supplied tables of the normalized differential cross section terms $(\rd \tilde{\sigma_i}/\rd E_{\text{jet}} \ \rd z)|_{z_1}$, computed for $\sqrt{s} = 8\TeV$ and jet radius $R = 0.5$. The cone algorithm used for the theoretical calculation does not introduce a systematic effect since there are no background or pileup sources in the theory. As described in Section~\ref{exanal}, we compare the data to sum of the $^1S_{0}^{8}$, $^3S_{1}^{8}$, and $^3P_{J}^{8}$ FJF functions for the LDME parameter sets from Bodwin, Chung, Kim, and Lee (BCKL)~\cite{bodwin}, Butenschoen and Kniehl (BK)~\cite{buten2}, and Chao, et al. (Chao)~\cite{chao}.  The LDME parameter sets for these three studies are derived from different selections of J/$\psi$ meson production measurements, e.g., the BK set includes electroproduction data and uses a lower J/$\psi$ meson $p_T$ limit than is used in the hadroproduction-only selection of the BCKL and Chao sets.  All groups report that their LDME sets yield J/$\psi$ meson differential cross sections that agree with the $\PJGy$ meson production data on which the extractions were based.

\section{Comparison of  data with FJF total cross section predictions}

\begin{figure*}[ht]
\centering
\includegraphics[width=.49\textwidth]{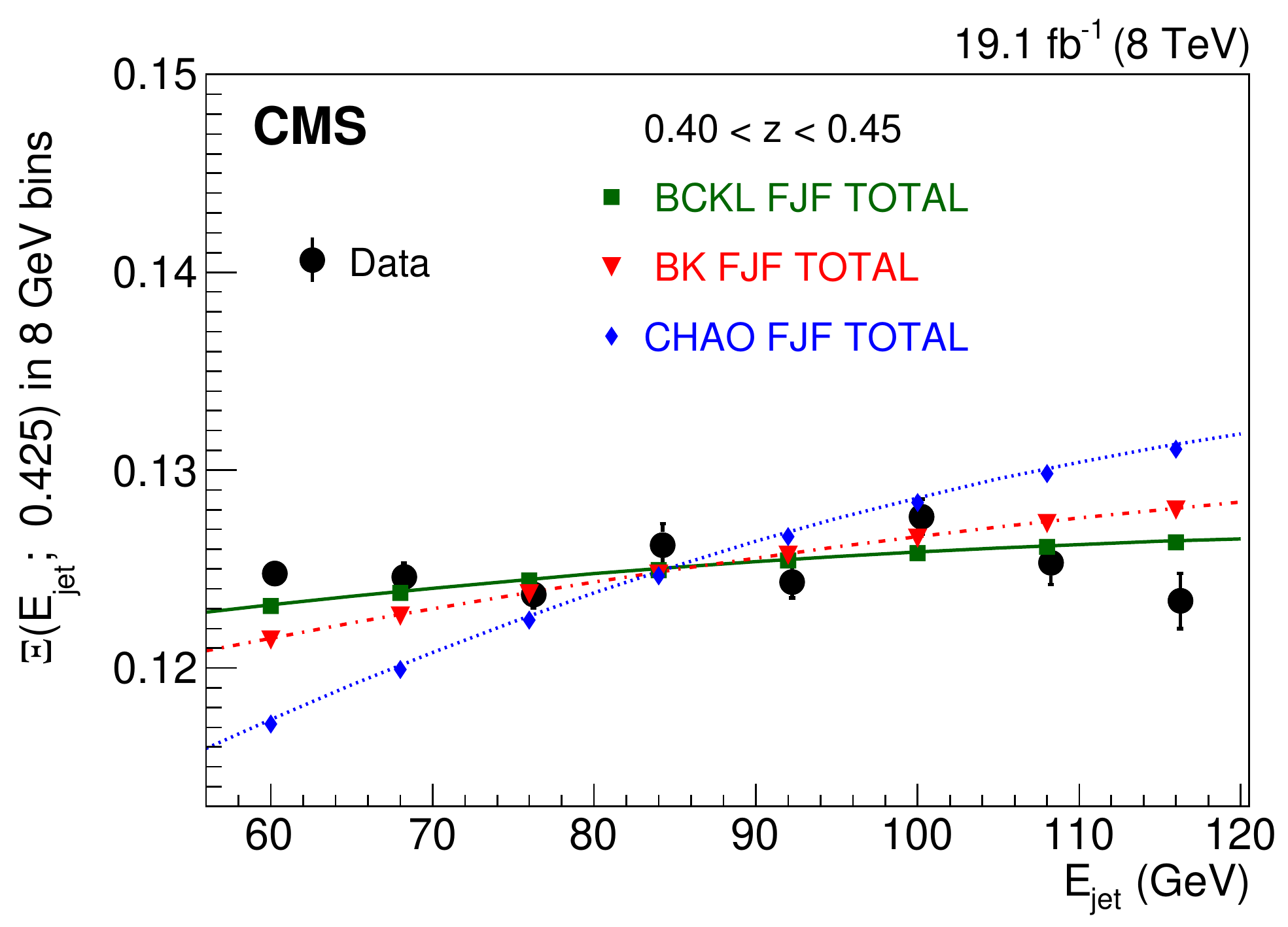} \includegraphics[width=.49\textwidth]{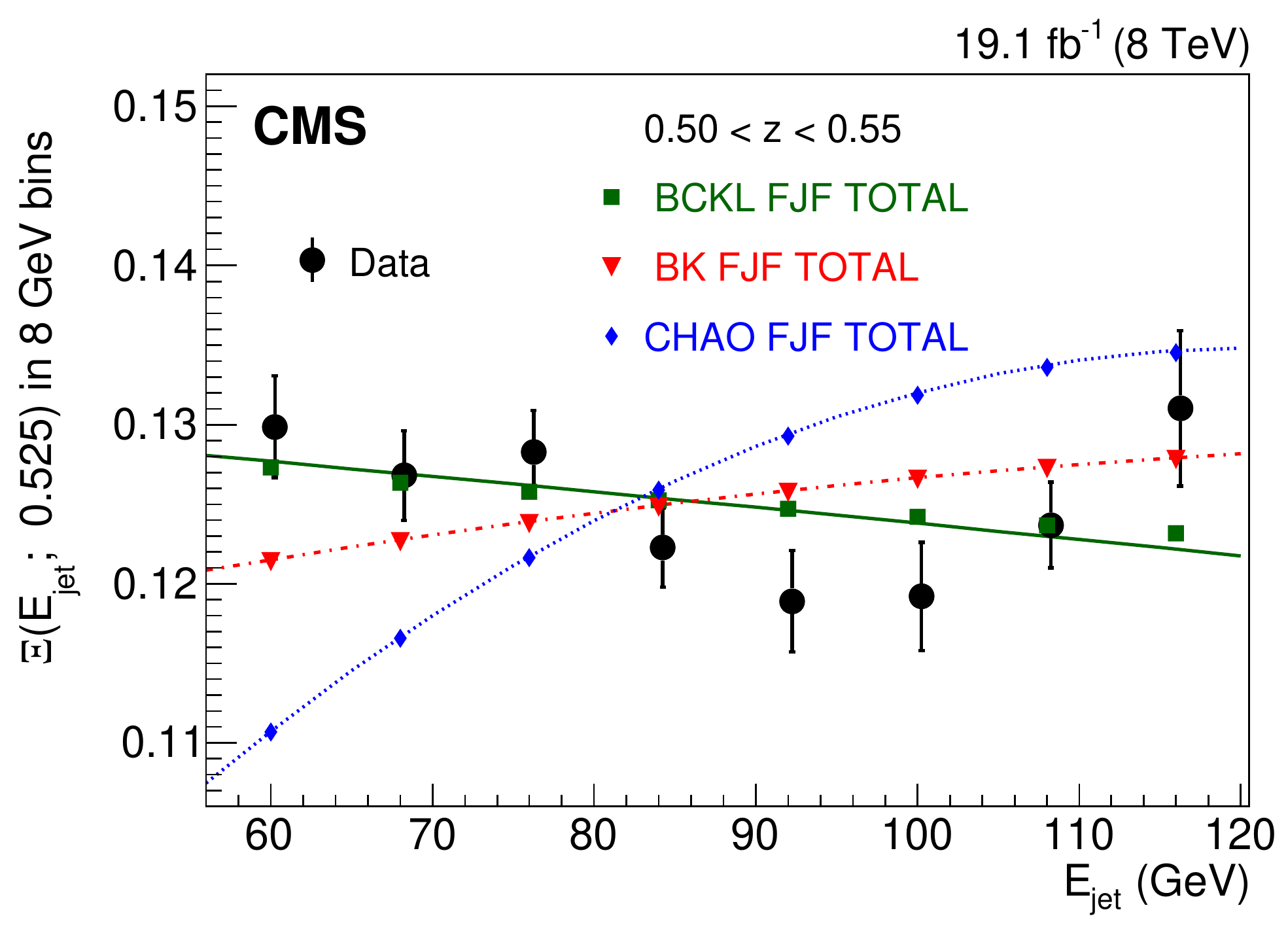} \includegraphics[width=.49\textwidth]{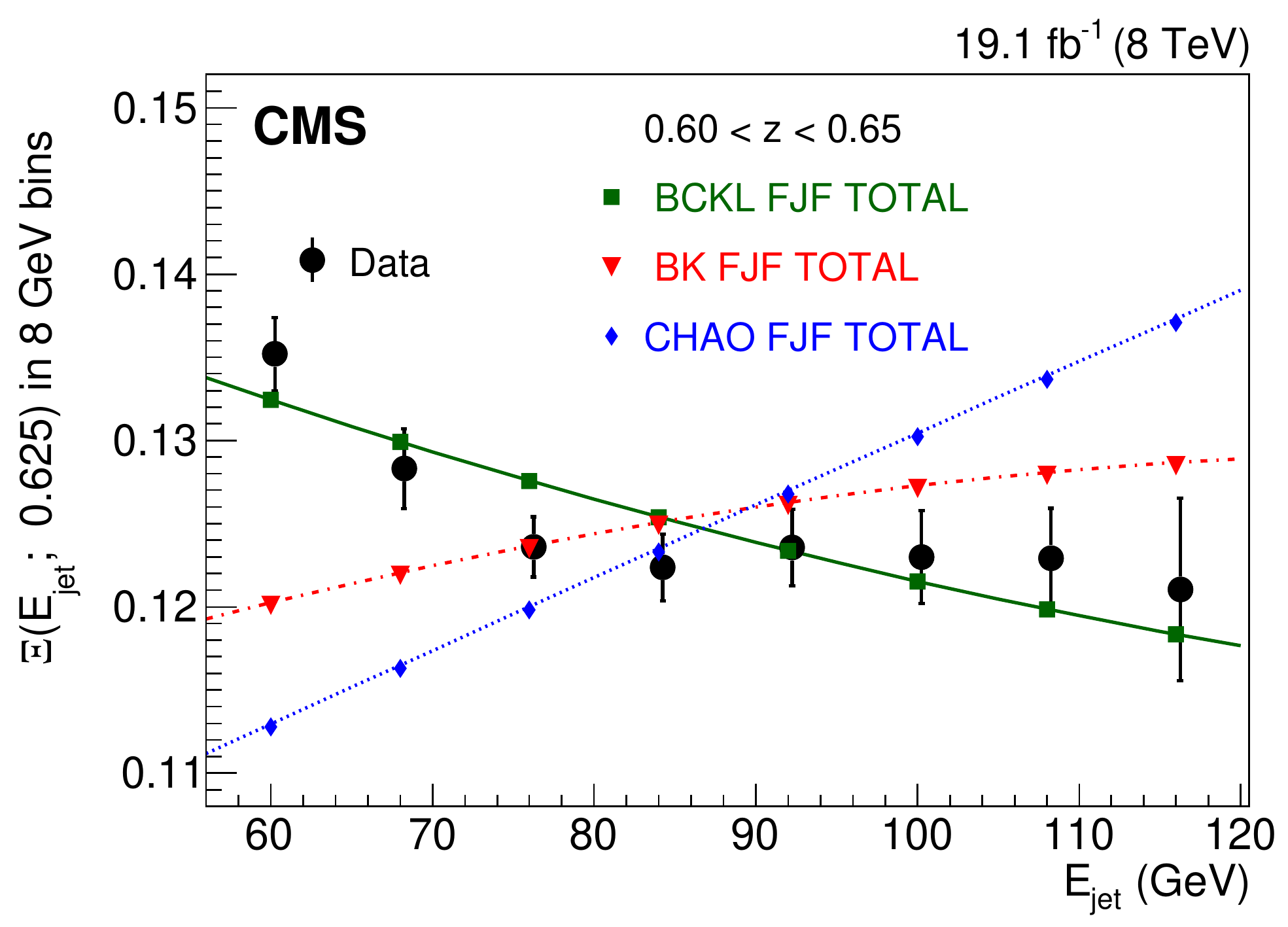} 
\caption{Comparison of $\Xi(E_{\text{jet}}; \ z_1)$ versus $E_{\text{jet}}$ from data with FJF predictions of the total differential cross section, each normalized to unit area, for the BCKL (squares), BK (inverted triangles), and Chao (diamonds) LDME parameter sets. The three $z_1$ ranges are (upper left) $z_1$ = 0.425; (upper right) $z_1$ = 0.525; (lower) $z_1$ = 0.625.  The curves show the detailed energy dependence of the predictions.  The vertical bars on the data points are the quadrature sum of the statistical and systematic uncertainties.\label{tot}}
\end{figure*}

In this section we compare the data with the prediction for the FJF total differential cross section from each of the three LDME sets.  Figure~\ref{tot} shows the normalized jet energy distributions for the data and the FJF total cross section predictions for each LDME set at each of the three $z_1$ values used in the analysis.  The uncertainties in the data include the statistical and systematic components added in quadrature.  For each $z_1$, the bin-averaged FJF values are used to calculate the $\chi^2$ for the comparison of the FJF total differential cross section prediction to the data.  The LDME calculations from Refs.~\cite{bodwin,buten2,chao} have normalization uncertainties, as shown in Ref.~\cite{ira2}.  The FJF calculations give the ratio of the cross section in a small-$z$ region to the cross section integral over a wide $z$ range.  The uncertainty in the predicted FJF values due to the theory normalization uncertainty is almost completely correlated for the numerator and denominator of the ratio.  The resulting theoretical uncertainty is negligible compared to the experimental uncertainty in the $\Xi(E_{\text{jet}}; z_1)$ values.  We therefore ignore it in computing the $\chi^2$ values to match data and theory.  The $\chi^2$ value and the associated $p$-value for comparison of data to each LDME set are given in Table~\ref{tottab}.  An a priori decision was made that a model prediction is an acceptable match to the data only if the $\chi^2$ p-value is larger than 0.1\% for seven degrees of freedom.  Otherwise, we say that the model does not match the data.  For all three $z_1$ ranges, the FJF predictions using the BCKL LDME parameters match data.  Neither the BK nor the Chao LDME parameter sets describe these jet + constituent $\PJGy$ data for any $z_1$ value.  

The observation that these new data on $\PJGy$ meson production as constituents of jets match the FJF predictions for the BCKL LDME parameter set and reject two others validates the FJF approach to treating jets as a major source of $\PJGy$ production in the gluon-rich central region in $\Pp\Pp$ interactions.  It also demonstrates that the BCKL LDME parameter set can describe new features of $\PJGy$ hadronic production at large $\PT^{\PJGy}$.  The BCKL LDME parameters were developed from a completely different data set than these $\PJGy$ + jet data, so there is no a priori reason to expect them to have predicted these measurements.  The BCKL parameters are known to predict small $\PJGy$ polarization~\cite{bodwin}, in agreement with experiment~\cite{prl99,cmspol} for the range of $\PT^{\PJGy}$ values selected in this analysis (10-40 GeV).  Because this analysis studies only high-$\PT$ $\PJGy$ meson production and shows that the BCKL LDME parameters describe the process and other sets do not, it suggests a tension between high-$\PT$ $\PJGy$ results and global charmonium studies~\cite{jpl}.

\begin{table*}[h]
\centering
\topcaption{The $\chi^2$ value and the associated $p$-value (in parentheses) for 7 degrees of freedom from the comparison of the data and the predictions for the total FJF cross section shape at $z_1$ = 0.425, 0.525, and 0.625, using the BCKL, the BK, and the Chao LDME parameter sets.  }
\begin{tabular}{lccc}
 & 0.425  & 0.525 & 0.625  \\ \hline
BCKL &  22.2 (0.23\%)  & 11.0 (14\%) & 10.7 (15\%) \\
BK & 59.6 (${<}0.001$\%)  & 60.1 (${<}0.001$\%)  & 64.0 (${<}0.001$\%) \\
Chao & 267 (${<}0.001$\%)  & 96 (${<}0.001$\%) & 164 (${<}0.001$\%) \\ 
\end{tabular}
\label{tottab}
\end{table*}

\section{Total fraction of \texorpdfstring{\PJGy}{J/psi} mesons from jets}\label{prod}

In this section, we determine whether jets are the major source of prompt energetic $\PJGy$ mesons ($E_{\PJGy} >15$\GeV) in the central region ($\abs{\eta_{\text{jet}}}<1$).  Here, $E_{\text{jet}}$ refers to the measured jet energy before unfolding.  As shown in Fig.~\ref{jetfrac1} (\cmsLeft), for events with a $\PJGy$ meson and only one observed jet, ($84.0\pm0.1$)\% of the $\PJGy$ candidates are within $\DR <0.5$ of that jet.  This is consistent with jets being the dominant source of $\PJGy$ production in this kinematic range when there is at least one observed jet in the event.  However, events with one or more observed jets having $\PT^{\text{jet}} >25$\GeV account for only ($44.9\pm0.1$)\% of the prompt $\PJGy$ meson sample.

To understand the source of $\PJGy$ meson events with no jets passing the $\PT^{\text{jet}}>25\GeV$ requirement, termed zero-jet events, we note that a jet that has a constituent $\PJGy$ meson can fail the $\PT^{\text{jet}}$ threshold even though the $\PJGy$ meson is observed.  For instance, when the $\PT^{\text{jet}}$ threshold is raised from 25 to 30\GeV, the fraction of zero-jet events with an identified $\PJGy$ meson increases from 55 to 65\%.  For one-jet events in data with $\PT^{\text{jet}}$ thresholds of 30, 35, and 40\GeV, the observed jet is found within $\DR <0.5$ of the $\PJGy$ meson in the event ($84.0\pm0.2$)\% of the time, i.e., the probability of a jet having a constituent $\PJGy$ meson is independent of $\PT^{\text{jet}}$.  Only jets with $E_{\text{jet}}>44\GeV$ pass the $\PT^{\text{jet}} > 25\GeV$ requirement with 100\% efficiency over the range $0<\abs{\eta_{\textrm{jet}}}<1$. Jets having $E_{\text{jet}}<44\GeV$ can contain observed $\PJGy$ mesons with $E_{\PJGy}>15\GeV$, but some of these jets will not pass the $\PT^{\text{jet}} >25\GeV$ requirement.

In order to correct for this effect, we fit the $E_{\text{jet}}$ distribution for jets containing a $\PJGy$ meson to the sum of two exponential functions in the range $44<E_{\text{jet}}<150\GeV$.  We use the fit to extrapolate the number of jets containing a $\PJGy$ meson to lower $E_{\text{jet}}$ values, in order to estimate the number of jets with a constituent $\PJGy$ meson that would be present in the lower-energy region for full $\PT$ acceptance.  Jet reconstruction efficiency corrections are not applied at this stage.    The FJF model is valid for $z <0.8$~\cite{ira}.   In the data, only ($1.3\pm0.1$)\% of jets having a constituent $\PJGy$ meson have $z >0.8$; we truncate the model at $z=0.8$, setting a limit of $E_{\text{jet}} >19\GeV$ for the extrapolation.  Some jets in the $E_{\text{jet}} = $ 25--44\GeV range have sufficiently large polar angles to pass the $\PT^{\text{jet}}>25\GeV$ requirement.  These are subtracted from the extrapolation to avoid double counting.  The number of jets from extrapolation in each 1\GeV wide jet energy bin $i$ is corrected for the jet reconstruction efficiency $\epsilon_i$ to predict the total number $N_i$ of jets with energy $E_i$.

In order to contribute to the data sample, a jet with energy $E_i$ must produce a $\PJGy$ meson with energy $E_j$.  The probability $P_j$ for the $\PJGy$ meson to have energy $E_j$ is taken from the results of this analysis, normalized to unity for 55 bins covering the range $15< E_{\PJGy} <70$\GeV.  The total number $A_i$ of jets with energy $E_i$ that contain a $\PJGy$ meson with energy fraction $z_{ij} =  E_{j}/E_{i}$ in the range 0.3--0.8 is
\begin{linenomath}
\begin{equation}
A_i = N_{i} \ \sum_{j=1}^{55} P_j \ w(z_{ij}). \label{addeqn}
\end{equation}
\end{linenomath}
The function $w(z_{ij})$ is the probability that a jet of energy $E_i$ will contain a $\PJGy$ meson having energy $E_j$.  To proceed, we need a specific model for the jet and $\PJGy$ kinematics.  We use the jet fragmentation model in Ref.~\cite{ira} for $E_{\text{jet}} =50\GeV$.  The probability is zero for $z > 0.8$.  The model predicts that ($43\pm3\stat$)\% of the $\PJGy$ mesons should be accompanied by zero observed jets, compared to 55\% found in the data.

There are systematic uncertainties in this result.  In a private communication, the authors of Ref.~\cite{ira} also provided a $z$ probability calculation for $E_{\textrm{jet}} =20\GeV$.  The model prediction for the number of zero-jet events using the 20\GeV $z$ probability calculation differs by 3\% from the 50\GeV result.  This difference is taken as the systematic uncertainty in the $z$ fragmentation probability.   The uncertainty in the MC prediction of the low-energy jet efficiency is 13\%.  We also made a closure test by using the model to predict the number of observed jets lost when the jet $\PT$ threshold was raised from 25 to 40\GeV.  The model prediction agrees with the actual number of lost jets to within (3.5 $\pm$ 0.1)\%.  However, there is a jet energy dependence in the matching between the data and the prediction.  Extrapolating the bin-by-bin jet energy dependence of that difference into the 19--44\GeV range, the closure study gives a 7\% systematic uncertainty in the predicted number of zero-jet events having jet energies less than 44\GeV.  Adding the systematic uncertainties in quadrature, the predicted fraction of zero-jet events with a $\PJGy$ meson as a constituent of a jet with $\PT^{\text{jet}} <25\GeV$ is ($43\pm3\stat\pm7\syst$)\%.  

If we apply this reasoning to results from Section~\ref{assn}, the small peak in the range $2.5< \DR <3.4$ in Fig.~\ref{jetfrac1} (\cmsLeft) is actually the recoil jet in a dijet pair for which the parent jet of the $\PJGy$ meson was not observed.  This increases the fraction of $\PJGy$ mesons that are constituents of a jet in the one-jet sample from ($84.0\pm0.1$)\% to ($94.3\pm0.1$)\%.  With this interpretation, and the results from Section~\ref{assn}, we find that the one- and two-jet fractions for a jet to have a constituent $\PJGy$ meson are both essentially 94\%.  The overall fraction of $\PJGy$ mesons that come from jets is, then, 0.94$\cdot$45\% = 42\% from events with one or more observed jets, plus 43\% from the zero-jet sample.  While the zero-jet model is simple, it passes an experimental closure test.  Also, it follows the trend of the data as the jet $\PT$ requirement is raised in steps from 25 to 40\GeV.  Using it, we conclude that ($85\pm3\stat\pm7\syst$)\% of the $\PJGy$ mesons within our kinematic acceptance, $E_{\PJGy}>15\GeV$ and $\abs{y_{\PJGy}}<1$, are constituents of jets with $E_{\text{jet}}>19\GeV$ and $\abs{\eta_{\text{jet}}}<1$. 

\section{Summary}

The first analysis has been presented comparing data for prompt $\PJGy$ mesons produced as constituents of central gluonic jets with a theoretical analysis based on the fragmenting jet function (FJF) approach. The term prompt means that the $\PJGy$ meson is consistent with originating from the primary vertex.  In the FJF model, the jet fragments into a $\PQc\PAQc$ system in an angular momentum state and quark color configuration $^{2S+1}L_{J}^{n}$,  plus other hadrons. Here, $S$, $L$, and $J$ are the spin, orbital, and total angular momentum quantum numbers of the $\PQc\PAQc$ system and $n$ indicates a color-singlet ($n$ = 1) or color-octet ($n$ = 8) configuration.  The FJF analysis uses the nonrelativistic quantum chromodynamics (NRQCD) approach to compute the cross section for the formation of a $\PJGy$ meson from the $\PQc\PAQc$ system for four specific $S$, $J$, $L$, and $n$ configurations: $^{1}S_{0}^{8}$, $^3S_{1}^{8}$, $^{3}S_{1}^{1}$, and $^3P_{J}^{8}$.

The data were collected by the CMS Collaboration in proton-proton collisions at $\sqrt{s} = 8\TeV$, corresponding to an integrated luminosity of $19.1 \fbinv$.  The kinematic selections for the analysis are $E_{\PJGy}>15\GeV$, $\abs{y_{\PJGy}}<1$, $\pt^{\text{jet}}>25\GeV$, and $\abs{\eta_{\text{jet}}}<1$.  In $z$ ranges 0.40--0.45, 0.50--0.55, and 0.60--0.65, where $z$ is the $\PJGy$ meson fraction of the jet energy, the shape of the masured differential cross section as a function of $E_{\text{jet}}$ for $\PJGy$ meson production as a jet constituent is compared to the FJF prediction for this quantity, using three different long-distance matrix element (LDME) parameter sets.  The FJF predictions using the Bodwin, Chung, Kim, and Lee (BCKL)~\cite{bodwin} LDME parameters match the data for all three $z$ ranges.  In contrast, the FJF predictions for the LDME parameter sets from Butenschoen and Kniehl (BK)~\cite{buten2} and Chao, \etal, (Chao)~\cite{chao} disagree with the data for all three $z$ ranges.  This establishes the ability of the FJF analysis to describe $\PJGy$ meson production from central gluonic jets and the ability of this kind of experiment to distinguish among different sets of LDME parameters, all of which describe inclusive $\PJGy$ meson production for their choice of data.  The BCKL LDME set, constructed using inclusive hadronic production data with $\PT^{\PJGy}>10\GeV$, not only describes the production of high-$\PT$ $\PJGy$ mesons as constituents of jets but also predicts small $\PJGy$ meson polarization.

When a jet is observed in an event, the fraction of $\PJGy$ mesons that are jet constituents is ($94.2 \pm 0.1$)\%, averaged over one- and two-jet events.  Using a simple model to estimate the fraction of $\PJGy$ mesons that are constituents of jets that fail the analysis $\PT^{\text{jet}}$ requirement, we find that jets are the source of ($85\pm3\stat\pm7\syst$)\% of the $\PJGy$ mesons produced in the kinematic region probed in this study.  Interpreting the results in the framework of the FJF model, jet fragmentation accounts for almost all prompt $\PJGy$ mesons produced at large $\PT^{\PJGy}$.  The data are consistent with an NRQCD treatment of the FJF process using the BCKL parameter set. 

\begin{acknowledgments}

We thank Ira Rothstein, Matthew Baumgart, and Prashant Shrivastava for their help in creating the tables of the predictions from their fragmenting jet function model to compare with our data.  We thank Nora Brambilla for editorial input and theoretical guidance.

We congratulate our colleagues in the CERN accelerator departments for the excellent performance of the LHC and thank the technical and administrative staffs at CERN and at other CMS institutes for their contributions to the success of the CMS effort. In addition, we gratefully acknowledge the computing centers and personnel of the Worldwide LHC Computing Grid for delivering so effectively the computing infrastructure essential to our analyses. Finally, we acknowledge the enduring support for the construction and operation of the LHC and the CMS detector provided by the following funding agencies: BMBWF and FWF (Austria); FNRS and FWO (Belgium); CNPq, CAPES, FAPERJ, FAPERGS, and FAPESP (Brazil); MES (Bulgaria); CERN; CAS, MoST, and NSFC (China); COLCIENCIAS (Colombia); MSES and CSF (Croatia); RPF (Cyprus); SENESCYT (Ecuador); MoER, ERC IUT, and ERDF (Estonia); Academy of Finland, MEC, and HIP (Finland); CEA and CNRS/IN2P3 (France); BMBF, DFG, and HGF (Germany); GSRT (Greece); NKFIA (Hungary); DAE and DST (India); IPM (Iran); SFI (Ireland); INFN (Italy); MSIP and NRF (Republic of Korea); MES (Latvia); LAS (Lithuania); MOE and UM (Malaysia); BUAP, CINVESTAV, CONACYT, LNS, SEP, and UASLP-FAI (Mexico); MOS (Montenegro); MBIE (New Zealand); PAEC (Pakistan); MSHE and NSC (Poland); FCT (Portugal); JINR (Dubna); MON, RosAtom, RAS, RFBR, and NRC KI (Russia); MESTD (Serbia); SEIDI, CPAN, PCTI, and FEDER (Spain); MOSTR (Sri Lanka); Swiss Funding Agencies (Switzerland); MST (Taipei); ThEPCenter, IPST, STAR, and NSTDA (Thailand); TUBITAK and TAEK (Turkey); NASU and SFFR (Ukraine); STFC (United Kingdom); DOE and NSF (USA). 

\hyphenation{Rachada-pisek} Individuals have received support from the Marie-Curie program and the European Research Council and Horizon 2020 Grant, contract No. 675440 (European Union); the Leventis Foundation; the A.P.\ Sloan Foundation; the Alexander von Humboldt Foundation; the Belgian Federal Science Policy Office; the Fonds pour la Formation \`a la Recherche dans l'Industrie et dans l'Agriculture (FRIA-Belgium); the Agentschap voor Innovatie door Wetenschap en Technologie (IWT-Belgium); the F.R.S.-FNRS and FWO (Belgium) under the ``Excellence of Science -- EOS" -- be.h project n.\ 30820817; the Ministry of Education, Youth and Sports (MEYS) of the Czech Republic; the Lend\"ulet (``Momentum") Program and the J\'anos Bolyai Research Scholarship of the Hungarian Academy of Sciences, the New National Excellence Program \'UNKP, the NKFIA research grants 123842, 123959, 124845, 124850, and 125105 (Hungary); the Council of Science and Industrial Research, India; the HOMING PLUS program of the Foundation for Polish Science, cofinanced from European Union, Regional Development Fund, the Mobility Plus program of the Ministry of Science and Higher Education, the National Science Center (Poland), contracts Harmonia 2014/14/M/ST2/00428, Opus 2014/13/B/ST2/02543, 2014/15/B/ST2/03998, and 2015/19/B/ST2/02861, Sonata-bis 2012/07/E/ST2/01406; the National Priorities Research Program by Qatar National Research Fund; the Programa Estatal de Fomento de la Investigaci{\'o}n Cient{\'i}fica y T{\'e}cnica de Excelencia Mar\'{\i}a de Maeztu, grant MDM-2015-0509 and the Programa Severo Ochoa del Principado de Asturias; the Thalis and Aristeia programs cofinanced by EU-ESF and the Greek NSRF; the Rachadapisek Sompot Fund for Postdoctoral Fellowship, Chulalongkorn University and the Chulalongkorn Academic into Its 2nd Century Project Advancement Project (Thailand); the Welch Foundation, contract C-1845; and the Weston Havens Foundation (USA). 
\end{acknowledgments}

\bibliography{auto_generated}

\cleardoublepage \appendix\section{The CMS Collaboration \label{app:collab}}\begin{sloppypar}\hyphenpenalty=5000\widowpenalty=500\clubpenalty=5000\input{BPH-15-003-authorlist.tex}\end{sloppypar}
\end{document}

%% file: BPH-15-003-authorlist.tex
\vskip\cmsinstskip
\textbf{Yerevan Physics Institute, Yerevan, Armenia}\\*[0pt]
A.M.~Sirunyan$^{\textrm{\dag}}$, A.~Tumasyan
\vskip\cmsinstskip
\textbf{Institut f\"{u}r Hochenergiephysik, Wien, Austria}\\*[0pt]
W.~Adam, F.~Ambrogi, T.~Bergauer, J.~Brandstetter, M.~Dragicevic, J.~Er\"{o}, A.~Escalante~Del~Valle, M.~Flechl, R.~Fr\"{u}hwirth\cmsAuthorMark{1}, M.~Jeitler\cmsAuthorMark{1}, N.~Krammer, I.~Kr\"{a}tschmer, D.~Liko, T.~Madlener, I.~Mikulec, N.~Rad, J.~Schieck\cmsAuthorMark{1}, R.~Sch\"{o}fbeck, M.~Spanring, D.~Spitzbart, W.~Waltenberger, C.-E.~Wulz\cmsAuthorMark{1}, M.~Zarucki
% ...
\vskip\cmsinstskip
\dag: Deceased\\
1:  Also at Vienna University of Technology, Vienna, Austria\\
% ...
